\documentclass[superscriptaddress,amssymb,amsmath,nobibnotes,aps,prd,eqsecnum,showpacs,
nofootinbib]{revtex4}
\usepackage{graphics,color}
\usepackage{bm}
\usepackage{graphicx}
\usepackage{amsmath}
\usepackage{amssymb}
\usepackage{enumitem}
\usepackage{epstopdf}
\usepackage[colorlinks=true]{hyperref}
\allowdisplaybreaks
\usepackage[dvipsnames]{xcolor}

\def\be{\begin{equation}}
\def\ee{\end{equation}}
\def\bea{\begin{eqnarray}}
\def\eea{\end{eqnarray}}

\def\nn{\nonumber \\}

\newcommand{\e}{\mathrm{e}}

\begin{document}

\title{Is exponential gravity a viable description for the whole cosmological history?}
\author{Sergei D. Odintsov}
\email{odintsov@ice.csic.es}
\affiliation{Institut de Ci\`{e}ncies de l'Espai, ICE/CSIC-IEEC, Campus UAB, Carrer de Can Magrans s/n, 08193 Bellaterra (Barcelona), Spain}
\affiliation{Instituci\'o Catalana de Recerca i Estudis Avan\c{c}ats (ICREA), Barcelona, Spain}
\author{Diego S\'aez-Chill\'on G\'omez}
\email{saez@ice.csic.es}
\affiliation{Institut de Ci\`{e}ncies de l'Espai, ICE/CSIC-IEEC, Campus UAB, Carrer de Can Magrans s/n, 08193 Bellaterra (Barcelona), Spain}
\author{German~S.~Sharov}
\email{sharov.gs@tversu.ru}
\affiliation{Tver state university 170002, Sadovyj per. 35, Tver, Russia}


\begin{abstract}
Here we analysed a particular type of $F(R)$ gravity, the so-called exponential gravity which includes an exponential function of the Ricci scalar in the action. Such term represents a correction to the usual Hilbert-Einstein action. By using Supernovae Ia, Barionic Acoustic Oscillations, Cosmic Microwave Background and $H(z)$ data, the free parameters of the model are well constrained. The results show that such corrections to General Relativity become important at cosmological scales and at late-times, providing an alternative to the dark energy problem. In addition, the fits do not  determine any significant difference statistically with respect to the $\Lambda$CDM model. Finally, such model is extended to include the inflationary epoch in the same gravitational Lagrangian. As shown in the paper, the additional terms can reproduce the inflationary epoch and satisfy the constraints from Planck data.
\end{abstract}


\maketitle

\section{Introduction}
\label{Intro}

Over the last decade, the study of some modifications of General Relativity have drawn a lot of attention, particularly in the framework of cosmology as an attempt to provide a more natural explanation to the accelerating expansion at early times (inflation) and at late times (dark energy epoch). In this sense, the most simple and natural extension of GR arises as the generalisation of the Hilbert-Einstein action by assuming a non-linear function of the Ricci scalar, what is commonly called $f(R)$ gravity (for a review see \cite{Reviews}). Other extensions include curvature invariants as the Gauss-Bonnet gravity \cite{Gauss-Bonnet} or generalizations of the so-called Teleparallel gravity, an equivalent theory to GR constructed as a gauge theory of the translation group leading to a null-curvature theory with non null torsion (see Ref.~\cite{Teleparallel}). Nevertheless, $f(R)$ gravities have been by far the most analysed extension of GR over the last years, also due to its motivation on more fundamental theories as string theory \cite{Nojiri:2003rz}. This extensive study has provided a very deep knowledge and comprehension of this type of theories, whose field equations turn out forth order differential equations instead of second order as in GR. Nevertheless, $f(R)$ gravities can be easily reduced to a type of scalar-tensor theory, i.e. $f(R)$ gravity basically implies the appearance of an extra scalar mode \cite{scalar-tensor}. As every theory with extra propagating modes, this may imply the existence of ghosts. Fortunately, this is not the case in $f(R)$ gravities. However, the extra scalar mode may imply violations and deformations of well known and tested predictions of GR. In order to avoid large corrections at scales where GR is very well tested, $f(R)$ gravities can hide such extra mode through a mechanism known as chameleon mechanism, proposed initially in the framework of scalar-tensor theories \cite{Khoury:2003aq}, but rapidly extended to $f(R)$ gravities \cite{HuSawicki07,Nojiri:2007as}. \\

In addition, the versatility of $f(R)$ gravities allows to reconstruct any cosmological solution with the suitable evolution \cite{Capozziello2002}.  Then, late-time acceleration may arise in a natural way as a consequence of the gravitational theory instead of being the aftermath of any extra unknown field. Moreover, simultaneously $f(R)$ gravities may contribute to the compensation of the large value predicted by quantum field theories for the vacuum energy density, and particularly may play an essential role in the framework of the so-called unimodular gravity theories. On this regard, $f(R)$ gravity scenarios as an alternative to the $\Lambda$CDM-cosmology are interesting and attractive, since they are able to describe simultaneously the early-time inflation as well as the late-time  acceleration in the expansion of our Universe \cite{scalar-tensor,Nojiri:2007as,unifying,CognolaENOSZ08}. Particularly, some of the most promising inflationary models are constructed within the $f(R)$ gravity scenario, since some of these models can easily reproduce slow-roll inflation by mimicking a cosmological constant at early times and then decaying, leading to a power spectrum for scalar perturbations nearly invariant and a negligible scalar to tensor ratio, coinciding with the last data released by the Planck collaboration \cite{Planck-Inflation}. This is the case for instance of Starobinsky inflation \cite{Starobinsky:1980te}, a quadratic Lagrangian on the Ricci scalar that predicts the correct values for the spectral index and the scalar to tensor ratio. Actually, some analysis suggest that any deviation from Starobinsky inflation should be small enough to avoid deviations from its well established predictions \cite{delaCruz-Dombriz:2016bjj}. Keeping this in mind, over the last years some efforts have been focused on the attempt to unify inflation and dark energy epoch in the framework of $f(R)$ gravities, and particularly within the so-called {\it viable} $f(R)$ gravity models \cite{Nojiri:2007as}. As mentioned above, these viable models accomplish the well known local tests, where the scalar mode acquires a large mass through the chameleon mechanism avoiding large corrections with respect to GR. Hence, the local tests or the Solar System tests for viable $f(R)$ theories include correct Newtonian and post-Newtonian limit\cite{HuSawicki07,Nojiri:2007as}. In addition, this type of models are capable of reproducing the correct late-time acceleration, in general by simulating an effective cosmological constant that becomes important at late-times, while $\Lambda$CDM behaviour is recovered at high redshift. Moreover, these models provide good fits when compared with observational data, being almost indistinguishable from $\Lambda$CDM \cite{delaCruz-Dombriz:2015tye}. However, viable $f(R)$ gravities contain a type of future cosmological singularity, the so-called {\it sudden} singularity, a consequence directly related to the mass of the scalar field that avoids corrections at local scales \cite{Frolov:2008uf}, although such singularity occurs in the future when the right parameters are set and can be avoided by adding some extra terms. Moreover, some extensions of such models are also capable of reproducing inflation at early times, when tends asymptotically to a power Lagrangian, leading to a Starobinsky-like inflation keeping the right predictions \cite{Nojiri:2007as}. \\

In this paper, we focus on the analysis of a type of viable $f(R)$ models that reproduces late-time acceleration by mimicking a cosmological constant but where corrections may have some distinguible effects. This class of $f(R)$ models are given by a negative exponential of the Ricci scalar in the action, which turns out negligible at large redshifts but becomes important at late-times, an effect easily controlled with a free parameter related to current Hubble parameter. Exponential gravity has been previously analysed in Refs.~\cite{Tsujikawa:2007xu,CognolaENOSZ08,ElizaldeNOSZ11,Linder2009} as a reliable alternative to other viable $f(R)$ gravities, since GR results are recovered at local scales but reproduce dark energy behaviour at cosmological ones. In addition, previous analysis has shown the existence of an asymptotically stable de Sitter solution in such exponential Lagrangians, leading to an approximated $\Lambda$CDM behaviour at the present time \cite{ElizaldeNOSZ11,Linder2009}.  Moreover, some recent analysis of such type of exponential gravities suggest that observational constraints can be well satisfied from the cosmological point of view, in such a way that $f(R)$ gravity and $\Lambda$CDM model turn out nearly indistinguishable, as suggested by previous analysis \cite{BambaGL2010,YangLeeLG2010,ChenGLLZ14}. In addition, exponential gravity can be extended to cover the inflationary stage as well. To do so, an additional exponential is considered in the gravitational action becoming important at large curvature when the inflationary period occurs, and turning out negligible as curvature decreases \cite{CognolaENOSZ08,ElizaldeNOSZ11}. Hence, in this paper we analyse such type of $f(R)$ gravities, firstly by fitting the free parameters of the model by using data from Type Ia supernovae, baryon acoustic oscillations (BAO), estimations of the Hubble parameter $H(z)$ and parameters of the cosmic microwave background radiation (CMB) \cite{Union21SNe,Eisen05,Planck13,Planck15}, and also considering different approaches. Then, we analyse how the full gravitational Lagrangian can cover also the inflationary epoch, obtaining the the spectral index for scalar perturbations and the tensor-to-scalar ratio.\\
%
%
%
%
%
%
%
%
%

The paper is organised as follows: section \ref{Dynamics} reviews the basics of $f(R)$ gravities, while section \ref{sect3} is devoted to introduce the exponential $f(R)$ gravity model and its dynamical equations. In Sect.~\ref{Data} the observational data considered in the paper is shown, this includes Union 2.1 observations of Type Ia supernovae, BAO effects, the latest measurements of the Hubble parameter  $H(z)$ and CMB parameters.  In Sect.~\ref{Analysis1} we estimate
the constraints on  the exponential $F(R)$ model from the aforementioned data. In Sect.~\ref{Inflation} we investigate the variant of the exponential model with inflation terms in the Lagrangian. Finally section \ref{conclusions} is devoted to the conclusions of the paper.

\section{F(R) gravity}
\label{Dynamics}

Modified $F(R)$ gravities are described by the following generalisation of the Einstein-Hilbert action \cite{Capozziello2002}
 \begin{equation}
  S = \frac1{2\kappa^2}\int d^4x \sqrt{-g}\,F(R)  + S^{m}.
 \label{Act1}\end{equation}
where $\kappa^2=8\pi G$ and $S^{m}$ is the matter action. Einstein General Relativity is very well understood and tested at many scales, so that one should assume the action (\ref{Act1}) to contain slightly deviations from GR, such that we can rewrite the action in the following way:
\be
F(R)=R+f(R)\ .
\label{dsg1}
\ee
Here, the function $f(R)$ accounts for the gravitational modifications and should become negligible at scales where GR is well tested. By varying the action (\ref{Act1}) with respect to the metric tensor $g_{\mu\nu}$, the field equations are obtained,
\begin{equation}
F_R R_{\mu\nu}-\frac F2
g_{\mu\nu}+\big(g_{\mu\nu}g^{\alpha\beta}\nabla_\alpha\nabla_\beta-\nabla_\mu\nabla_\nu\big)F_R
=\kappa^2T_{\mu\nu}\ ,
 \label{eq1}\end{equation}
 where $R$ and $R_{\mu\nu}$ are the Ricci scalar and Ricci tensor respectively, whereas $F_R\equiv F'(R)$ and $T_{\mu\nu}$ is the energy-momentum tensor of matter. Note that $F(R)$ field equations are fourth order in comparison to the second order of General Relativity. However, the action (\ref{Act1}) hides an additional scalar mode, such that can be expressed as the Lagrangian of a type of scalar-tensor theory as follows:
 \be
 S = \frac1{2\kappa^2}\int d^4x \sqrt{-g}\,\left[\phi R-V(\phi)\right]  + S^{m}\ ,
 \label{Act2}
\ee
where the following relations are found:
\be
\phi=F_{R}\ , \qquad V(\phi)=RF_R-F\ .
\label{dsg2}
\ee
Hence, in order to avoid large deviations from GR, this additional degree of freedom should be hidden at the appropriate scale, a mechanism commonly known as the chameleon mechanism \cite{Khoury:2003aq}.  In this sense, some $F(R)$ actions which accomplish this requirement have been proposed in the literature \cite{HuSawicki07,Nojiri:2007as}, particularly some of them with the form of a negative exponential, the type of Lagrangians we are exploring in this manuscript. Nevertheless, let us first analyse the general properties of $F(R)$ gravities, and in particular in the cosmology framework. By assuming a spatially-flat Friedman-Lema\^itre-Robertson-Walker (FLRW) space-time with the metric
 $$ds^2 =-dt^2 + a^2(t)\,d\mathbf{x}^ 2,$$
where $a(t)$ is the scale factor of the universe, $c=1$, the Ricci scalar is expressed as:
 \begin{equation}
 R = 6 (2H^2 + \dot H )
 \label{RH}\end{equation}
Here, the Hubble parameter is defined as usual by $H=\dot a/a$, where the dot denotes
derivatives with respect to the cosmic time. By assuming an energy-momentum tensor $T^{\mu}_{\nu}
= \mbox{diag}\,(-\rho,p,p,p)$ as a perfect fluid, where $\rho$ and $p$ are the matter
energy density and pressure, the field equations (\ref{eq1}) turn out
\cite{Capozziello2002}
 \begin{eqnarray}
&& \qquad H^2F_R+\frac16(F-RF_R)+H\dot F_R=\frac13 \kappa^2\rho,  \nonumber \\
&& \!\!\!\!\!\! (2\dot H+3H^2)\,F_R+\frac12(F-RF_R)+2H\dot F_R+\ddot F_R=-\kappa^2 p\ .
\label{dsg3}
 \end{eqnarray}
While the divergence of the field equations lead to the energy conservation equation $\nabla^\mu T_{\mu\nu}=0$, which in a FLRW metric becomes:
\be
\dot{\rho}+3H(\rho+p)=0\ .
\label{dsg4}
\ee
The FLRW equations (\ref{dsg3}) can be expressed in terms of other independent variables instead of the the cosmic time for convenience. Here we use the number of e-folds, given by $x=\log a=-\log(z+1)$ with $a(t_0)=1$ at the present time $t_0$, to express the above equations in the form of a dynamical system as follows
 \begin{eqnarray}
\frac{dH}{dx}&=&\frac{R}{6H}-2H, \nonumber \\
\frac{dR}{dx}&=&\frac1{F_{RR}}\bigg(\frac{\kappa^2\rho}{3H^2}-F_R+\frac{RF_R-F}{6H^2}\bigg),
 \nonumber\\
 \frac{d\rho}{dx}&=&-3(\rho+p). \label{dynam} \end{eqnarray}
 Here $F_{RR}\equiv F''(R)$ are derivatives with respect to $x$ and we have used the Ricci scalar definition Eq.~(\ref{RH}) and the continuity equation (\ref{dsg4}). Hence,  the analysis of the above system can provide all the information about the dynamics produced by a particular action $F(R)$.

\section{Exponential gravity}
\label{sect3}

Let us now introduce the type of exponential $F(R)$ gravity, we are considering in this manuscript \cite{CognolaENOSZ08,ElizaldeNOSZ11,Linder2009,BambaGL2010,YangLeeLG2010,ChenGLLZ14,Tsujikawa:2007xu}
 \begin{equation}
F(R)=R-2\Lambda\bigg[1-\exp\Big(-\frac R{R_0}\Big)\bigg]=
R-2\Lambda\bigg[1-\exp\Big(-\beta\frac{R}{2\Lambda}\Big)\bigg]\ .
 \label{FR1}\end{equation}
The model contains just two free parameters $\Lambda$ and $R_0$, which may be expressed in a more convenient way as $R_0=2\Lambda/\beta$, where $\beta$ is dimensionless \cite{BambaGL2010,YangLeeLG2010,ChenGLLZ14}
 $$\beta=2\Lambda/R_0.$$
 Note that in principle the model (\ref{FR1}) can well describe the universe evolution for $z<10^4$, including
the recombination epoch, the  matter-dominated era and the late-time acceleration. This is true as far as $\beta\geq0$, since the exponential becomes negligible and the action (\ref{FR1}) recovers the usual $\Lambda$CDM model at large redshifts, where the curvature becomes much larger than $\Lambda$. In Sect.~\ref{Inflation} below we will also consider corrections to this model such that early-time inflation is also described.\\

Here we focus on the epoch for $0\le z\le10^4$, when the content of the universe includes
 pressureless (non-relativistic) matter  and radiation (relativistic particles):
 $\rho=\rho_m+\rho_r$, such that the continuity equation (\ref{dsg4}) can be solved and yields
  \begin{equation}
 \rho=\rho_m^0a^{-3}+ \rho_r^0a^{-4},
 \label{rho}\end{equation}
 where $\rho_m^0$ and $\rho_r^0$ are the present time values of these components, which can be normalised over the critical density as follows:
 \be
 \Omega_{i}=\frac{\rho_{0i}}{\frac{3}{\kappa^2}H_0^2}\ .
 \label{dsg5}
 \ee
 Here $H_0$ is the Hubble parameter today. Let us first explore the behaviour of model (\ref{FR1}) during the early universe, when the curvature becomes large $R\to\infty$ as $z\to\infty$ (or for $z\ge10^4$ in
practical applications). Then, the model (\ref{FR1}) transforms into the  $\Lambda$CDM
model with $F(R)=R-2\Lambda$, so the solutions of the system (\ref{dynam}) tend
asymptotically to $\Lambda$CDM at large redshifts, leading to:
 \begin{equation}
 \frac{H^2}{(H^{\Lambda CDM}_0)^2}=\Omega_m^{\Lambda CDM} \big(a^{-3}+ X^{\Lambda CDM}a^{-4}\big)+\Omega_\Lambda^{\Lambda CDM},\qquad
 \frac{R}{2\Lambda}=2+\frac{\Omega_m^{\Lambda CDM}}{2\Omega_\Lambda^{\Lambda CDM}}a^{-3}, \qquad a\to0.
  \label{asymLCDM}\end{equation}
 Here the index ``$\Lambda CDM$'' refers to quantities calculated within the $\Lambda$CDM model, where  $\Omega_\Lambda^{\Lambda CDM}=\frac\Lambda{3(H^{\Lambda CDM}_0)^2}$ and $H^{\Lambda CDM}_0$ the Hubble parameter today as predicted by the $\Lambda$CDM model, while $X=\Omega_r/\Omega_m$. However, despite the model (\ref{FR1}) recovers $\Lambda$CDM at large redshifts, late-time evolution deviates from $\Lambda$CDM, such that the above quantities as measured today $t=t_0$ would differ from $\Lambda$CDM unless initial conditions are fixed at $z=0$, which is not the case of our paper. Note that other viable $f(R)$ models shows a similar behaviour when are analysed asymptotically \cite{HuSawicki07}. Hence, we have that in general:
 $$
 H_0\ne H^{\Lambda CDM}_0, \qquad \Omega_m^0\ne \Omega_m^{\Lambda CDM}\ ,
 $$
where we have denoted by $0$ those magnitudes measured today as predicted by our model (\ref{FR1}). Nevertheless, we can connect both models through the relation of the physical matter density \cite{HuSawicki07}
 \begin{equation}
 \Omega_m^0H_0^2=\Omega_m^{\Lambda CDM}(H^{\Lambda CDM}_0)^2=\frac{\kappa^2}3\rho_m(t_0),
  \label{H0Omm}\end{equation}
As will be shown below, this remark is important when performing the fitting analysis for the observable parameters in Sect.~\ref{Data}. Moreover, note that the first FLRW equation (\ref{dsg3}) for the $\Lambda$CDM model is a constraint equation which evaluated at $t=t_0$ can be expressed as follows:
\be
\Omega_m^{\Lambda CDM}+\Omega^{\Lambda CDM}_\Lambda=1\ .
\label{dsg6}
\ee
 This expression is very well known in standard cosmology when GR is assumed but breaks down when other gravitational actions beyond GR are considered, as $F(R)$ gravity. In such case, the first FLRW equation becomes a dynamical equation, since it contains second derivatives of the Hubble parameter. By evaluating the FLRW equation in (\ref{dsg3}) at $z=0$, the above equation can be expressed as:
 \begin{equation}
 \Omega_m^0+\Omega_\Lambda^0=1-\Omega_{f(R_0)}^0\ .
  \label{OmmL}\end{equation}
Note that here we have defined $\Omega_{\Lambda}^0=\frac{\Lambda}{3H_0^2}$, which refers to the cosmological constant term in the action (\ref{FR1}), while $\Omega_{f(R)}^0$ includes the exponential function in (\ref{FR1}). The smaller $\Omega_{f(R)}^0$ is, the closer our model is to $\Lambda$CDM  at the present time, where the expression (\ref{OmmL}) is evaluated. Nevertheless, note that our model recovers $\Lambda$CDM asymptotically at high redshifts ($z>10$) such that the differences among the relative densities $\Omega_i(z)$ become negligible in both models at high redshifts.\\

Let us now for convenience in the calculations, introduce the following dimensionless
variables  
\begin{equation}
E=\frac{H}{H_0^{\Lambda CDM}}, 
\qquad {\cal R}=\frac{R}{2\Lambda},
   \label{Er}\end{equation}
 Hence, the gravitational action (\ref{FR1}) becomes:
 $$F(R)=2\Lambda({\cal R}-1+e^{-\beta {\cal R}}),$$
  while the system of equations (\ref{dynam}) takes the form
\begin{eqnarray}
\frac{dE}{dx}&=&\Omega_\Lambda^{\Lambda CDM}\frac{{\cal R}}{E}-2E, \label{eqE2} \\   
\frac{d{\cal R}}{dx}&=&\frac{e^{\beta{\cal R}}}{\beta^2}\bigg[\Omega_m^{\Lambda
CDM}\frac{a^{-3}+ X^{\Lambda CDM}a^{-4}} {E^2}-1+\beta e^{-\beta {\cal
R}}+\Omega_\Lambda^{\Lambda CDM}\frac{1-(1+\beta {\cal R})\,e^{-\beta {\cal
R}}}{E^2}\bigg].
  \label{eqr2}\end{eqnarray}
Recall  that the variable $x=\log a=-\log(z+1)$ refers to the number of e-folds. This
system can be solved numerically by setting the appropriate initial conditions. As
naturally for the model (\ref{FR1}), we assume initial conditions that match
$\Lambda$CDM model at a particular high redshift:
 \begin{equation}
 E^2(x_i)=\Omega_m^{\Lambda CDM} \big(e^{-3x_i}+ X^{\Lambda CDM}e^{-4x_i}\big)+\Omega_\Lambda^{\Lambda CDM},\qquad
  {\cal R}(x_i)=2+\frac{\Omega_m^{\Lambda CDM}}{2\Omega_\Lambda^{\Lambda CDM}}e^{-3x_i}\ ,
  \label{initial}\end{equation}
which corresponds to the $\Lambda$CDM asymptotic solution (\ref{asymLCDM}) at an initial
redshift  $z_i$, or alternatively at $x_i$. The value of $x_i$ is determined by assuming
the following condition:
 \be e^{-\beta {\cal R}(x_i)}=\varepsilon
\quad\Longleftrightarrow \quad x_i=\frac13\log\frac{\beta\Omega_m^{\Lambda
CDM}}{2\Omega_\Lambda^{\Lambda CDM}(\log\varepsilon^{-1}-2\beta)}, \label{dsg7} \ee
 where $\varepsilon$ is a small number in the range $10^{-10}<\varepsilon<10^{-7}$, such that our model mimics the $\Lambda$CDM
solution (\ref{asymLCDM}) at $x<x_i$, and the corresponding solutions practically do not depend on $\varepsilon$ or $x_i$ for all $x$.\\

Alternatively, we can also use the following variable \cite{HuSawicki07,ElizaldeNOSZ11,BambaGL2010,YangLeeLG2010,ChenGLLZ14}
\begin{equation}
 y_H=\frac{3H^2}{\kappa^2\rho_m^0}- a^{-3}- X^{\Lambda CDM}a^{-4}
  \label{yH}\end{equation}
 Then, the equation for $H$ in (\ref{dynam}) can be rewritten as follows
\begin{equation}
 \frac{dy_H}{dx}=-4y_H+2\frac{\Omega_\Lambda^{\Lambda CDM}}{\Omega_m^{\Lambda CDM}}{\cal R}-e^{-3x},\qquad
  y_H(x_i)=\frac{\Omega_\Lambda^{\Lambda CDM}}{\Omega_m^{\Lambda CDM}}\ ,
 \label{eqyH}\end{equation}
where the second expression corresponds to the initial condition (\ref{initial}). The advantage of the function $y_H$ and its derivative (\ref{eqyH}) lies on their finiteness at $z\to\infty$
($x\to-\infty$). However, the same does not apply for the equation (\ref{eqr2}), since $y_R=3(\kappa^2\rho_m^0)^{-1}R- 3a^{-3}$ is not finite for every redshift. Hence, numerical integration of the system Eq.~(\ref{eqyH}) or the corresponding second order differential equation for $y_H$ owns similar difficulties as the system (\ref{eqE2}), (\ref{eqr2}).

\section{Observational Data}
\label{Data}

Let us now present the data we are using here to fit the free parameters of our model. Besides the late-time evolution data from SNe Ia, BAO and $H(z)$, we are also considering the CMB parameters. Then, to do so, we have to include radiation in our equations, or in other words, assuming equation (\ref{initial}), or alternatively eq.~(\ref{eqyH}). As our model mimics $\Lambda$CDM at high redshifts, we can reduce the number of free parameters by fixing the radiation-matter ratio as provided by Planck \cite{Planck15}:
\be
X=\frac{\Omega_r}{\Omega_m}=2.9656\cdot10^{-4}\ .
\ee
Hence, our model contains 4 free parameters (\ref{FR1}):
\be
\beta, \quad \Omega_m^{\Lambda CDM}, \quad \Omega_\Lambda^{\Lambda CDM} \quad \text{and} \quad H_0^{\Lambda CDM}\ .
\label{FreeParam}
\ee
Remind that the Hubble parameter differs from the true Hubble constant $H_0=H_0^{\Lambda CDM}E\big|_{z=0}$, as well as the density parameters $\Omega_m^0H_0^2=\Omega_m^{\Lambda CDM}(H^{\Lambda CDM}_0)^2$. Nevertheless, the Hubble constant $H_0^{\Lambda CDM}$ can be considered as a  nuisance parameter, so that can be marginalized for all fits.\\

Here, we use the catalogue provided by Union 2.1 data, which contains 580 points  from Type Ia Supernovae (SNe Ia) \cite{Union21SNe}. BAO data described in Table \ref{TBAO}, Refs.~\cite{Gazta09}-\cite{Aubourg14}. We also use 30 estimations of the Hubble parameter $H(z)$ measured from differential ages of galaxies and summarised in Table \ref{TH}, \cite{Simon05}-\cite{Moresco16}. Finally, the CMB parameters are considered from the Planck mission \cite{Planck15}. In order to proceed with the analysis we use the technique of the minimum $\chi^2$, which establishes the best set of the parameters. To do so, we use a two dimensional grid, such that the free parameters (\ref{FreeParam}) are reduced to two, either by theoretical considerations or through marginalisation.

\subsection{Supernovae Ia data}

The  Union 2.1 compilation provides \cite{Union21SNe} $N_{SN}=580$ Sne Ia with their observed (estimated) distance moduli  $\mu_i=\mu_i^{obs}$ for redshifts $z_i$ in the interval $0 \leq z_i \leq 1.41$. In order to fit the free parameters of our model, we compare $\mu_i^{obs}$ with the theoretical value  $\mu^{th}(z_i) $, where the distance moduli is given by:
\begin{equation}
\mu (z)\equiv\mu^{th}(z) = 5 \log_{10} \frac{D_L(z)}{10\mbox{pc}}, \qquad D_L (z)= c
(1+z) \int_0^z \frac{d\tilde z}{H (\tilde z)} \label{mu}\ .
\end{equation}
Here $D_L (z)$ is the luminosity distance. The corresponding $\chi^2$ function is calculated by computing the differences between the SNe Ia observational data and the predictions of a particular model with
parameters $p_1,p_2,\dots$,
 \begin{equation}
\chi^2_{SN}(p_1,p_2,\dots)=\min\limits_{H_0} \sum_{i,j=1}^{N_{SN}}
 \Delta\mu_i\big(C_{SN}^{-1}\big)_{ij} \Delta\mu_j,
  \label{chiSN}\end{equation}
 where $\Delta\mu_i=\mu^{th}(z_i,p_1,\dots)-\mu^{obs}_i$, $C_{SN}$ is the $580\times580$
covariance matrix \cite{Union21SNe}. The marginalisation over the  nuisance parameter $H_0^{\Lambda}$ is widely described in the literature (see Refs.~\cite{Sharov16,Jimenez:2016sgs,Leanizbarrutia:2014xta}).

\subsection{BAO data}

Baryon acoustic oscillations (BAO) are obtained from galaxy clustering analysis and include measurements of two cosmological parameters \cite{Eisen05}
 \begin{equation}
 d_z(z)= \frac{r_s(z_d)}{D_V(z)},\qquad
  A(z) = \frac{H_0\sqrt{\Omega_m^0}}{cz}D_V(z),
  \label{dzAz} \end{equation}
where $r_s(z_d)$ is the sound horizon at the decoupling epoch and $D_V(z)$ is given by
$$
 D_V(z)=\bigg[\frac{cz D_L^2(z)}{(1+z)^2H(z)}\bigg]^{1/3}.
 $$
 The values (\ref{dzAz}) were estimated for redshifts $z=z_i$ (and redshift
ranges) of galaxies from a peak in the correlation function of the galaxy distribution at the comoving sound horizon scale $r_s(z_d)$, which corresponds to the decoupling of the photons $z_d$. In this paper we use the BAO data from Refs.~\cite{Gazta09}-\cite{Aubourg14} for the parameters (\ref{dzAz}), which provides $N_{BAO}=17$ data points for $d_z(z)$ and 7 data
points for $A(z)$, both shown in Table~\ref{TBAO}. We use the covariance matrices $C_{d}$ and $C_{A}$ for correlated data from Refs.~\cite{Percival09,BlakeBAO11} described in detail in Ref.~\cite{Sharov16}. So the $\chi^2$ function for the values (\ref{dzAz}) yields,
 \begin{equation}
 \chi^2_{BAO}(p_1,p_2,\dots)=\Delta d\cdot C_d^{-1}(\Delta d)^T+
\Delta { A}\cdot C_A^{-1}(\Delta { A})^T\ ,
  \label{chiB} \end{equation}
 where $\Delta d$ and $\Delta A$ are vector columns with  $\Delta d_i=d_z^{obs}(z_i)-d_z^{th}(z_i)$ and $\Delta A_i=A^{obs}(z_i)-A^{th}(z_i)$.

\begin{table}[th]
\centering
 {\begin{tabular}{||l|l|l|l|l|c|l||}
\hline
 $z$  & $d_z(z)$ &$\sigma_d$    & ${ A}(z)$ & $\sigma_A$  & Refs & Survey\\ \hline
 0.106& 0.336  & 0.015 & 0.526& 0.028& \cite{Beutler11}  & 6dFGS \\ \hline
 0.15 & 0.2232 & 0.0084& -    & -    & \cite{Ross14}& SDSS DR7  \\ \hline
 0.20 & 0.1905 & 0.0061& 0.488& 0.016& \cite{Percival09,BlakeBAO11}  & SDSS DR7 \\ \hline
 0.275& 0.1390 & 0.0037& -    & -    & \cite{Percival09}& SDSS DR7 \\ \hline
 0.278& 0.1394 & 0.0049& -    & -    & \cite{Kazin09}  &SDSS DR7 \\ \hline
 0.314& 0.1239 & 0.0033& -    & -    & \cite{BlakeBAO11}& SDSS LRG \\ \hline
 0.32 & 0.1181 & 0.0026& -    & -    & \cite{Anderson14} &BOSS DR11 \\ \hline
 0.35 & 0.1097 & 0.0036& 0.484& 0.016& \cite{Percival09,BlakeBAO11} &SDSS DR7 \\ \hline
 0.35 & 0.1126 & 0.0022& -    & -    & \cite{Padmanabhan12}   &SDSS DR7 \\ \hline
 0.35 & 0.1161 & 0.0146& -    & -    & \cite{ChuangW12}   &SDSS DR7 \\ \hline
 0.44 & 0.0916 & 0.0071& 0.474& 0.034& \cite{BlakeBAO11}& WiggleZ \\ \hline
 0.57 & 0.0739 & 0.0043& 0.436& 0.017& \cite{Chuang13}& SDSS DR9 \\ \hline
 0.57 & 0.0726 & 0.0014& -    & -    & \cite{Anderson14}& SDSS DR11 \\ \hline
 0.60 & 0.0726 & 0.0034& 0.442& 0.020& \cite{BlakeBAO11} & WiggleZ \\ \hline
 0.73 & 0.0592 & 0.0032& 0.424& 0.021& \cite{BlakeBAO11} &WiggleZ \\ \hline
 2.34 & 0.0320 & 0.0021& -& - & \cite{Delubac14} & BOSS DR11 \\ \hline
 2.36 & 0.0329 & 0.0017& -& - & \cite{Font-Ribera13} & BOSS DR11 \\  \hline
 \end{tabular}
 \caption{Values of $d_z(z)=r_s(z_d)/D_V(z)$ and $A(z)$ (\ref{dzAz})
with errors and references} \label{TBAO}}\end{table}

As pointed above, the Hubble parameter today $H_0$ as predicted by our model (\ref{FR1}) differs from the one predicted by the $\Lambda$CDM model $H_0^{\Lambda CDM}$, which is considered here as a free parameter. Both are related by the expression $H_0=H_0^{\Lambda CDM}E(t_0)$. However, the theoretical values of $d_z$ and $A$ (\ref{dzAz}) do not contain $H_0$, since the distances $D_L$  (\ref{mu}), $D_V$ and  $r_s(z_d)$ are proportional to
$H_0^{-1}$. In the expression $A(z)$ we can use the equivalence (\ref{H0Omm}) $H_0\sqrt{\Omega_m^0}=H_0^{\Lambda CDM}\sqrt{\Omega_m^{\Lambda CDM}}$.\\

All these considerations have to be carefully studied in order to choose the appropriate approach to calculate the
sound horizon $r_s(z_d)$ from different fitting formulae \cite{YangLeeLG2010,WangW2013,HuangWW2015,Aubourg14}. Here we are considering the following
simple fitting formula \cite{Sharov16}
 \begin{equation}
 r_s(z_d)=\frac{104.57\mbox{ Mpc}}h,\qquad
 h=\frac{H_0}{100\mbox{ km}/(\mbox{s}\cdot\mbox{Mpc})}\ ,
 \label{rsh}\end{equation}
 with explicit $h$ dependence. For the $\Lambda$CDM model, one obtains $(r_d\cdot h)_{fid}=104.57\pm1.44$ Mpc as the best fit (see Ref.~\cite{Sharov16} ). Other approaches give the same predictions as using
Eq.~(\ref{rsh}), see Ref.~\cite{Aubourg14}.

\subsection{$H(z)$ data}

The Hubble parameter parameter values $H$ at certain redshifts $z$ can be measured with
two methods: (1) extraction $H(z)$ from line-of-sight BAO data
\cite{Gazta09,Blake12,Busca12,ChuangW12,Chuang13,Anderson13,Anderson14,Oka13,Font-Ribera13,Delubac14}
and (2)  $H(z)$ estimations from differential ages $\Delta t$ of galaxies
\cite{Simon05,Stern10,Moresco12,Zhang12,Moresco15,Moresco16} via the following relation:
 $$ 
 H (z)= \frac{\dot{a}}{a} \simeq -\frac{1}{1+z}
\frac{\Delta z}{\Delta t}.
 $$ 

To avoid additional correlation with the BAO data from Table~\ref{TBAO}, we use in this
paper only $N_H=30$ values $H(z)$ estimated from differential ages of galaxies, shown in
Table~\ref{TH}. The theoretical values $H^{th}(z_i, p_1,\dots)$ naturally depend on
$H_0$. so the $\chi^2$ function is marginalized over $H_0$ \cite{SharovBPNC17}:
\begin{equation}
\tilde\chi^2_{H}= \sum_{i=1}^{N_H} \left[\frac{H^{obs}(z_i)-H^{th}(z_i,
p_j)}{\sigma_{H,i}}\right]^2,\qquad
\chi^2_{H}=\min\limits_{H_0}\tilde\chi^2_{H}.\label{chiH}
\end{equation}

\begin{table}[h]
\centering
 {\begin{tabular}{||l|l|l|c||l|l|l|c||}   \hline
 $z$  & $H(z)$ &$\sigma_H$  & Refs  &   $z$ & $H(z)$ & $\sigma_H$  & Refs\\ \hline
 0.070 & 69  & 19.6& \cite{Zhang12}  & 0.4783& 80.9& 9   & \cite{Moresco16}\\ \hline
 0.090 & 69  & 12  & \cite{Simon05}  & 0.480 & 97  & 62  & \cite{Stern10}  \\ \hline
 0.120 & 68.6& 26.2& \cite{Zhang12}  & 0.593 & 104 & 13  & \cite{Moresco12} \\ \hline
 0.170 & 83  & 8   & \cite{Simon05}  & 0.6797& 92  & 8   & \cite{Moresco12} \\ \hline
 0.1791& 75  & 4   & \cite{Moresco12}& 0.7812& 105 & 12  & \cite{Moresco12}\\ \hline
 0.1993& 75  & 5   & \cite{Moresco12}& 0.8754& 125 & 17  & \cite{Moresco12}\\ \hline
 0.200 & 72.9& 29.6& \cite{Zhang12}  & 0.880 & 90  & 40  & \cite{Stern10}  \\ \hline
 0.270 & 77  & 14  & \cite{Simon05}  & 0.900 & 117 & 23  & \cite{Simon05}  \\ \hline
 0.280 & 88.8& 36.6& \cite{Zhang12}  & 1.037 & 154 & 20  & \cite{Moresco12}\\ \hline
 0.3519& 83  & 14  & \cite{Moresco12}& 1.300 & 168 & 17  & \cite{Simon05}  \\ \hline
 0.3802& 83  & 13.5& \cite{Moresco16}& 1.363 & 160 & 33.6& \cite{Moresco15}\\ \hline
 0.400 & 95  & 17  & \cite{Simon05}  & 1.430 & 177 & 18  & \cite{Simon05}  \\ \hline
 0.4004& 77  & 10.2& \cite{Moresco16}& 1.530 & 140 & 14  & \cite{Simon05}  \\ \hline
 0.4247& 87.1& 11.2& \cite{Moresco16}& 1.750 & 202 & 40  & \cite{Simon05}  \\ \hline
 0.445 & 92.8& 12.9& \cite{Moresco16}& 1.965 &186.5& 50.4& \cite{Moresco15}\\ \hline
 \end{tabular}
\caption{Hubble parameter values $H(z)$ with errors $\sigma_H$ from
Refs.~\cite{Simon05,Stern10,Moresco12,Zhang12,Moresco15,Moresco16}}
 \label{TH}}\end{table}

\subsection{CMB data}
\label{CMBdata}

Unlike the described above SNe Ia, BAO and $H(z)$ data corresponding to the late-time
era $0<z\le2.36$, cosmological observations associated with CMB radiation
\cite{WangW2013,HuangWW2015,Aubourg14} include parameters at the photon-decoupling epoch
$z_*\simeq1090$ ($z_*=1089.90 \pm0.30$ \cite{Planck15}), particularly the comoving
sound horizon $r_s(z_*)$  and the transverse comoving distance
  \begin{equation}
 r_s(z)=\frac1{\sqrt{3}}\int_0^{1/(1+z)}\frac{da}
 {a^2H(a)\sqrt{1+\big[3\Omega_b^0/(4\Omega_r^0)\big]a}}\ , \quad D_M(z_*)=\frac {D_L(z_*)}{1+z_*} = c \int_0^{z_*}
\frac{d\tilde z}{H (\tilde z)}\ .
  \label{rs2}\end{equation}
In the present manuscript, we use the CMB parameters in the following form \cite{WangW2013,HuangWW2015}
 \begin{equation}
  \mathbf{x}=\big(R,\ell_A,\omega_b\big)=\bigg(\sqrt{\Omega_m^0}\frac{H_0D_M(z_*)}c,\,\frac{\pi
  D_M(z_*)}{r_s(z_*)},\,\Omega_b^0h^2\bigg)
 \label{CMB} \end{equation}
 with the estimations (distance priors) from Ref.~\cite{HuangWW2015}
  \begin{equation}
  R^{Pl}=1.7448\pm0.0054,\quad \ell_A^{Pl}=301.46\pm0.094,\quad\omega_b^{Pl}=0.0224\pm0.00017.
   \label{CMBpriors} \end{equation}
 Here $\Omega_b^0$ is the present time baryon fraction. The distance priors (\ref{CMBpriors})
with their errors $\sigma_i$ and the the covariance matrix
$$C_{CMB}=\|\tilde
C_{ij}\sigma_i\sigma_j\|,\qquad
 \tilde C=\left(\begin{array}{ccc} 1 & 0.53 & -0.73\\ 0.53 & 1 & -0.42\\ -0.73 & -0.42 & 1
\end{array} \right)$$
 were derived in Ref.~\cite{HuangWW2015} from the Planck collaboration data \cite{Planck15}
with free amplitude of the lensing power spectrum. For the value $z_*$ we use the
fitting formula from Refs.~\cite{WangW2013,HuangWW2015,HuSugiyama95}; the sound horizon
$r_s(z_*)$ is estimated from Eq.~(\ref{rs2}) as the correction $\Delta
r_s=\frac{dr_s}{dz} \Delta z$.\\

Hence, the $\chi^2$ function corresponding to the data (\ref{CMB}-\ref{CMBpriors}) is obtained as follows
  \begin{equation}
\chi^2_{CMB}=\min_{H_0,\omega_b}\tilde\chi^2_{CMB},\qquad
\tilde\chi^2_{CMB}=\Delta\mathbf{x}\cdot
C_{CMB}^{-1}\big(\Delta\mathbf{x}\big)^{T},\qquad \Delta
\mathbf{x}=\mathbf{x}-\mathbf{x}^{Pl}
 \label{chiCMB} \end{equation}
 which is minimised by marginalizing over the additional parameter $\omega_b=\Omega_b^0h^2$,
which should be considered as a nuisance parameter,  as well as over $H_0$ or
$H_0^{\Lambda CDM}$. However, for the joint analysis of $H(z)$ and CMB data, the
marginalisation over $H_0$ is calculated simultaneously:
  \begin{equation}
\chi^2_H+\chi^2_{CMB}=\min_{H_0}\big(\tilde\chi^2_{H}+\min_{\omega_b}\tilde\chi^2_{CMB}\big).
 \label{chiHCMB} \end{equation}
Let us now present the results for the $f(R)$ model considered here.

\section{Testing exponential $F(R)$ gravity}
\label{Analysis1}

By considering the SNe Ia,  $H(z)$, BAO and CMB data illustrated in the previous
section, the above exponential model is well constrained. Here, we calculate these
limitations and the best-fitted values of the parameters for the exponential $F(R)$
model  (\ref{FR1}). After marginalizing over $H_0$ (and over  $\omega_b$ for the CMB data in
$\chi^2_{CMB}$), the $F(R)$ model (\ref{FR1}) owns 3 free parameters: $\beta$,
$\Omega_m^{\Lambda CDM}$ and $\Omega_\Lambda^{\Lambda CDM}$.
Remind that they differ from  $\Omega_m^{0}$ and $\Omega_\Lambda^{0}$, these values are
connected $\Omega_{m}^{0}=\Omega_m^{\Lambda CDM}/E^2(0)$  and
$\Omega_{\Lambda}^{0}=\Omega_{\Lambda}^{\Lambda CDM}/E^2(0)$, as shown above in
eq.~(\ref{H0Omm}). Consequently, the sum
$\Omega_m^0+\Omega_\Lambda^0=1-\Omega_{f(R)}^{0}\neq 1$ as given in (\ref{OmmL}). The
sum (\ref{dsg6}) $ \Omega_m^{\Lambda CDM}+\Omega_\Lambda^{\Lambda CDM}$ is also not
equal 1 in general for the considered $F(R)$ model.

However, firstly let us assume the following condition:
 \begin{equation}
 \Omega_m^{\Lambda CDM}+\Omega_\Lambda^{\Lambda CDM}=1.
  \label{OmmL1}  \end{equation}

This means that the model is assumed to be closed to $\Lambda$CDM. 
 This assumption relaxes the difficulties to fit the free
parameters, since the free parameters of the model can be automatically reduced, leading
to 2 free parameters: $\beta$ and $\Omega_m^{\Lambda CDM}$.

\begin{figure}[th]
   \centerline{ \includegraphics[scale=0.66,trim=5mm 0mm 2mm -1mm]{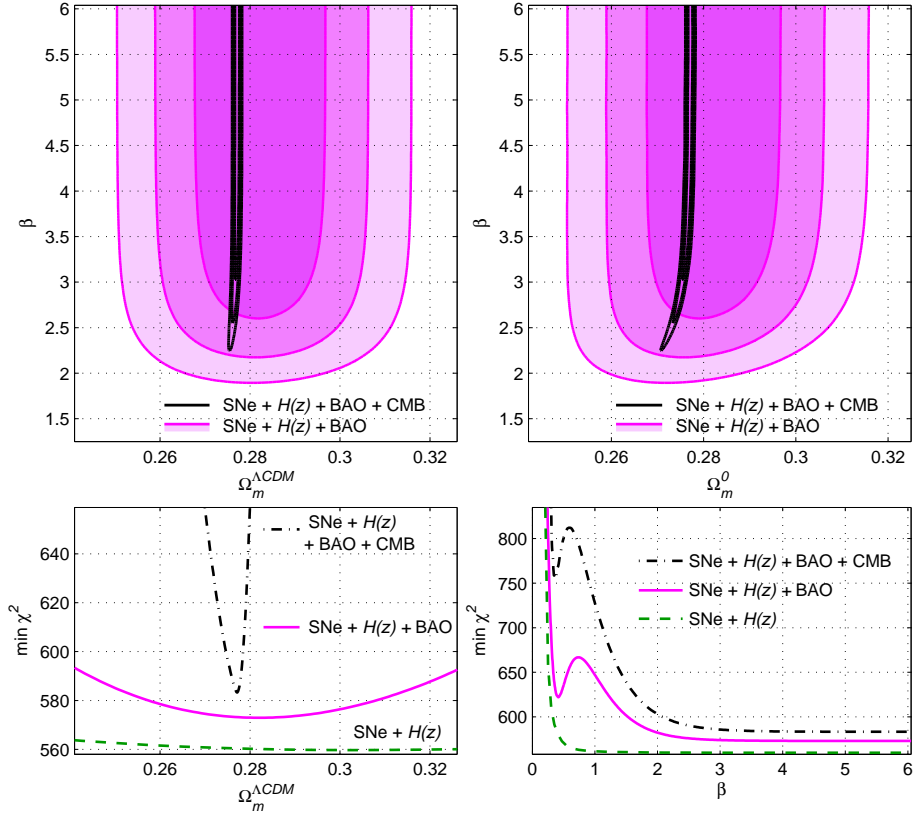}}
 \caption{Upper panels: contour plots for the free parameters of the exponential $F(R)$ model (\ref{FR1}) when assuming
 $\Omega_m^{\Lambda CDM}+\Omega_\Lambda^{\Lambda CDM}=1$,
 the left panel shows the $\Omega_m^{\Lambda CDM}-\beta$ plane while  the $\Omega_m^0-\beta$ plane is depicted in the right panel.
 Bottom panels: the one-dimensional dependencies of $\chi_{min}^2$ with respect to $\Omega_m^{\Lambda CDM}$ (left panel) and to $\beta$ (right panel).}
  \label{F1}
\end{figure}

The results are depicted in Fig.~\ref{F1} where the $1\sigma$, $2\sigma$ and $3\sigma$
regions are shown in the contour plots for the $\Omega_m^{\Lambda CDM}-\beta$ plane (the
top-left panel) and  for the $\Omega_m^0-\beta$ plane
 (the top-right panel). The magenta contours correspond to $\chi^2_{\Sigma3}=\chi^2_{SN}+\chi^2_H+\chi^2_{BAO}$ whereas the black lines
describe $\chi^2_{tot}=\chi^2_{SN}+\chi^2_H+\chi^2_{BAO}+\chi^2_{CMB}$.

   At each point in the $\Omega_m^{\Lambda CDM}-\beta$ plane, or in other
words, for given values of $\beta$, $\Omega_m^{\Lambda CDM}$ (and
$\Omega_\Lambda^{\Lambda CDM}=1-\Omega_m^{\Lambda CDM}$), the differential equations
(\ref{eqE2}), (\ref{eqr2}) are solved by assuming the $\Lambda$CDM model as the initial
conditions at high redshift (\ref{initial}). Then, once the solution $E(x)$ is
determined for each set of the free parameters, the  $\chi^2$ functions: $\chi^2_{SN}$
(\ref{chiSN}), $\chi^2_{BAO}$  (\ref{chiB}) are obtained.
 Furthermore, by considering the function $H(z)=H_0^{\Lambda CDM}E(z)$, we calculate then
the $\chi_{H}^2$ and the $\chi_{CMB}^2$ by marginalizing over $H_0^{\Lambda CDM}$ (and
over $\omega_b$ for $\chi^2_{CMB}$), or in other words keeping $H_0^{\Lambda CDM}$ as a
nuisance parameter, which avoids further bias on the results, so that we obtain the
optimal value  $H_0^{\Lambda CDM}$ (for these fixed $\beta$ and $\Omega_m^{\Lambda
CDM}$) and calculate the corresponding values $H_0=H_0^{\Lambda CDM}E(0)$ and
$\Omega_m^0=\Omega_m^{\Lambda CDM}(H^{\Lambda CDM}_0/H_0)^2$ from Eqs.~(\ref{Er}) and (\ref{H0Omm}), see Ref.~\cite{Bennett:2014tka}. \\

By following this procedure, we can calculate $\Omega_m^0$ at each point and draw the contour plots in the $\Omega_m^0-\beta$ plane, as shown in the top-right panel of Fig.~\ref{F1}. These
calculations were made separately for SNe${}+H(z)+{}$BAO data (the filled contours) and for SNe${}+H(z)+\mbox{BAO}+{}$CMB data (the black contours). The difference between the $\Omega_m^{\Lambda CDM}-\beta$ panel and the $\Omega_m^{\Lambda CDM}-\beta$ planel is clearly shown for small $\beta$, while in the limit $\beta\to \infty$ the model (\ref{FR1}) tends to the $\Lambda$CDM, where $\Omega_m^{\Lambda CDM}$ and $\Omega_m^0$ coincide.

In the bottom-left panel of Fig.~\ref{F1} the one-dimensional dependencies of
$\chi_{min}^2$ on $\Omega_m^{\Lambda CDM}$ are shown for
$\chi^2_{tot}=\chi^2_{SN}+\chi^2_H+\chi^2_{BAO}+\chi^2_{CMB}$ (the black dash-dotted
line),  for $\chi^2_{SN}+\chi^2_H+\chi^2_{BAO}$ (the solid magenta line) and for
$\chi^2_{SN}+\chi^2_H$ (the green dashed line). The latter is not depicted in the
upper panels, however for all 3 cases, the one-dimensional distributions $f(\Omega_m^{\Lambda CDM})$
are calculated from the corresponding two-dimensional matrices describing the contours in the top-left panel.
Hence, under the restriction (\ref{OmmL1}), the dependence of $\chi^2_{tot}$
with respect to the CMB data  (\ref{CMB}-\ref{CMBpriors}) on $\Omega_m^{\Lambda CDM}$ is very strong (unlike for the late-time SNe${}+H(z)+{}$BAO data $\chi^2_{\Sigma3}=\chi^2_{SN}+\chi^2_H+\chi^2_{BAO}$).

Similarly, the one-dimensional dependency of $\chi_{min}^2$ on $\beta$ (calculated from the two-dimensional matrices $\chi^2(\Omega_m^{\Lambda CDM}-\beta)$) are depicted at the bottom-right panel of Fig.~\ref{F1}. Under the restriction (\ref{OmmL1}) for all $\chi^2$ functions, the absolute minimum is
achieved at the limit $\beta\to\infty$, in other words, at the ``$\Lambda$CDM'' limit of the model (\ref{FR1}).

This conclusion may be illustrated in another way: the curves of the bottom-left panel of Fig.~\ref{F1} will coincide with the ones for the flat $\Lambda$CDM model, since these minima are achieved at large $\beta$,
where $\Omega_m^{\Lambda CDM}$ coincide with  $\Omega_m^0$.

While in the bottom-right panel of Fig.~\ref{F1}, one can see the unusual behaviour of the
one-dimensional distributions $\chi^2_{tot}(\beta)$ and $\chi^2_{\Sigma3}(\beta)$: these
functions have the local minima at $\beta\simeq0.4$. These minima are not shown in the
top panels of Fig.~\ref{F1}, because they lie beyond the $3\sigma$ confidence levels.
However, suitable level lines of $\chi^2=$const show these local minima as ``islands'' in the  $\Omega_m^{\Lambda CDM}-\beta$ or $\Omega_m^0-\beta$ planes. For instance, the corresponding coordinates or optimal values for $\chi^2_{\Sigma3}(\Omega_m^{\Lambda CDM},\beta)$ are $\Omega_m^{\Lambda CDM}\simeq0.252$, $\beta\simeq 0.415$.

\medskip

 Furthermore, let us now consider the model (\ref{FR1}) for its general case (\ref{OmmL})
beyond the restriction (\ref{OmmL1}) $ \Omega_m^{\Lambda CDM}+\Omega_\Lambda^{\Lambda
CDM}=1$.
 In this case
the model has 3 free parameters: $\beta$, $\Omega_m^{\Lambda CDM}$ and
$\Omega_\Lambda^{\Lambda CDM}$ (or alternatively $\Omega_m^{0}$ and
$\Omega_\Lambda^{0}$). So in order to calculate the corresponding $\chi^2$, the value of
the three parameters has to be given before solving numerically the system (\ref{eqE2}),
(\ref{eqr2}) with the initial conditions (\ref{initial}), as described above. In this
case every $\chi^2$ function (after marginalization over $H_0^{\Lambda CDM}$ and
$\omega_b$ for $\chi^2_H$ and $\chi^2_{CMB}$) will depend on $\beta$, $\Omega_m^{\Lambda
CDM}$ and $\Omega_\Lambda^{\Lambda CDM}$.

\begin{figure}[th]
   \centerline{ \includegraphics[scale=0.66,trim=5mm 0mm 2mm -1mm]{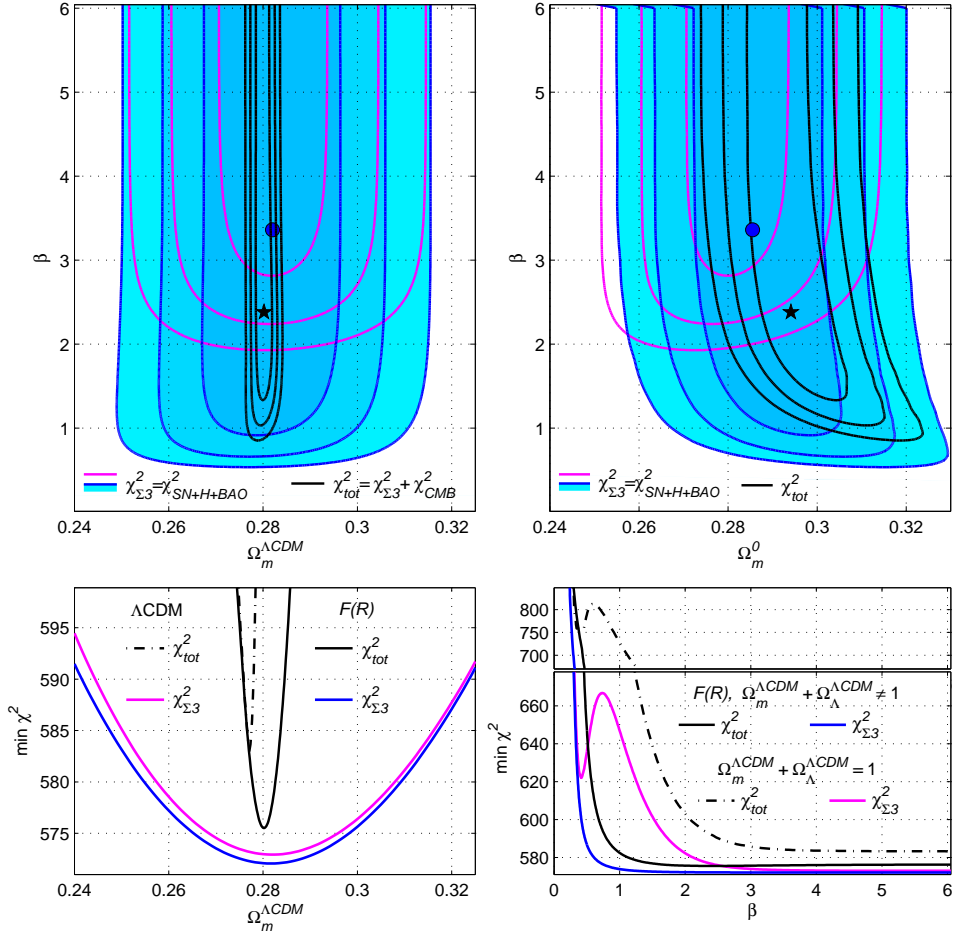}}
\caption{General case where $ \Omega_m^{\Lambda CDM}+\Omega_\Lambda^{\Lambda CDM}\ne1$.
Upper panels: contour plots for $\Omega_m^{\Lambda CDM}-\beta$ (left panel) and
$\Omega_m^0-\beta$ (right panel). Bottom plots: one-dimensional distributions for
$\chi^2_{tot}$ (solid black lines) and for $\chi^2_{\Sigma3}$ (solid blue lines and
filled contours) are compared with the ones from Fig.~\ref{F1} for $\chi^2_{\Sigma3}$
(magenta lines) and for $\chi^2_{tot}$ (thin black dash-dotted lines).}
  \label{F2}
\end{figure}

Hence, when we draw the contour plots for $\chi^2_{\Sigma3}$ and $\chi^2_{tot}$ in the
$\Omega_m^{\Lambda CDM}-\beta$ plane in the top-left panel of Fig.~\ref{F2}, we minimize
these functions over $\Omega_\Lambda^{\Lambda CDM}$ at each point of the plane. In other
words, we calculate $\chi^2_{min}(\Omega_m^{\Lambda
CDM},\beta)=\min\limits_{\Omega_\Lambda^{\Lambda CDM},H_0,\omega_b}\chi^2$ for
$\chi^2_{\Sigma3}$ and $\chi^2_{tot}$. 
 %

At each point of the $\Omega_m^{\Lambda CDM},\beta$ plane, the minima of the
$\chi^2_{\Sigma3}$ and $\chi^2_{tot}$ functions are calculated, also the optimal values
of the free parameters $\Omega_\Lambda^{\Lambda CDM}$, $H_0^{\Lambda CDM}$,
$H_0=H_0^{\Lambda CDM}E(0)$ and $\Omega_m^0=\Omega_m^{\Lambda CDM}/[E(0)]^2$ are
obtained. These values help us to draw the contour plots in the $\Omega_m^0,\beta$ plane
in the top-right panel of Fig.~\ref{F2}.

The same panels and notations of Fig.~\ref{F1} are used in Fig.~\ref{F2}, but the blue
contours corresponds to $\chi^2_{\Sigma3}(\Omega_m^{\Lambda CDM},\beta)$ in
Fig.~\ref{F2} while the blue lines refer to the one-dimensional distributions
$\chi^2_{\Sigma3\ min}(\Omega_m^{\Lambda CDM})$ and $\chi^2_{\Sigma3\ min}(\beta)$ in
the bottom panels of Fig.~\ref{F2}. In order to compare these results with those
obtained under the approximation (\ref{OmmL1}), the curves of Fig.~\ref{F1} are depicted
as well, denoted by magenta lines for $\chi^2_{\Sigma3\ min}$ and by thin black
dash-dotted lines for $\chi^2_{tot\ min}$ (in the bottom panels).

The black stars in Fig.~\ref{F2} denote the minimum points of the two-dimensional distributions $\chi^2_{tot}(\Omega_m^{\Lambda CDM},\beta)$ and $\chi^2_{tot}(\Omega_m^0,\beta)$.
Their coordinates (the optimal values of parameters) are tabulated in Table~\ref{Estim}. In the same way, the minimum points for $\chi^2_{\Sigma3}$  are shown as the blue circles.

The contour plots in Fig.~\ref{F2} demonstrate that the model (\ref{FR1}) in absence of the approximation (\ref{OmmL1}), and with the free
parameters $\beta$, $\Omega_m^{\Lambda CDM}$ and $\Omega_\Lambda^{\Lambda CDM}$ the regions of $1\sigma$,
$2\sigma$ or $3\sigma$ confidence level in the  $\Omega_m^{\Lambda CDM}-\beta$ plane are essentially enlarged in comparison with Fig.~\ref{F1}. Particularly, for $\chi^2_{\Sigma3}$ the $1\sigma$ domains (the  blue
filled contours) occupies the range $\beta>0.91$, whereas under the restriction (\ref{OmmL1}) (the magenta contours in Figs.~\ref{F1} and \ref{F2}) the range is $\beta>2.8$. For the joint function $\chi^2_{tot}$ (the black contours) these areas are larger in the $\beta$ direction and wider in the $\Omega_m^{\Lambda CDM}$ direction, especially for the parameter  $\Omega_m^0$, as shown in the top-right panel of Fig.~\ref{F2}.

These enlarged domains of suitable model parameters include the above mentioned
``islands'' or  local minima of $\chi^2_{\Sigma3}$ and $\chi^2_{tot}$ functions existed
under the restriction (\ref{OmmL1}). This effect is hidden in the top panels of
Fig.~\ref{F2} (it is beyond the $3\sigma$ confidence level), but it is shown in the
bottom-right panel, where the local minima at $\beta\simeq0.4$ from Fig.~\ref{F1} (for
the magenta and black dash-dotted lines) are naturally included in the general behaviour
of $\chi^2_{\Sigma3}$ (the blue line) and $\chi^2_{tot}$ (the solid black line). These
one-dimensional distributions determine the optimal values and  $1\sigma$ errors of the
parameter $\beta$; this information is included in Table~\ref{Estim}, where the absolute
minima of $\chi^2$ and the mean of the model parameters are provided.

The optimal values and $1\sigma$ errors for  $\Omega_m^{\Lambda CDM}$ in Table~\ref{Estim} are deduced from the one-dimensional distributions $\chi_{min}^2(\Omega_m^{\Lambda CDM})$. They are shown in the bottom-left panel of  Fig.~\ref{F2} as the solid black and blue curves in comparison with the corresponding plots for the case $\Omega_m^{\Lambda CDM}+\Omega_\Lambda^{\Lambda CDM}=1$ (the dash-dotted and magenta lines, they are taken from Fig.~\ref{F1}). The latter distributions coincide with the predictions of the $\Lambda$CDM model.

One can conclude from Table~\ref{Estim} (and Fig.~\ref{F2}) that the  absolute minima for the $F(R)$ model (\ref{FR1}) $\chi^2_{\Sigma3}\simeq 572.07$ and $\chi^2_{tot}\simeq 575.51$ are smaller than the ones for $\Lambda$CDM model (572.93 and 583.24 respectively). Such difference lies on the existence of degrees of freedom for the model (\ref{FR1}).

 \begin{table}[ht]
 \centering
 {\begin{tabular}{||l|c||c|c|c|c|l||}  \hline
 Model  & data & $\Omega_m^{\Lambda CDM}$& $\Omega_m^0$& $\Omega_\Lambda^{\Lambda CDM}$ & $\beta$ &$ \min\chi^2/d.o.f$  \\ \hline
$\!F(R)\;$(\ref{FR1})$\!$& $\chi^2_{\Sigma3}$& $0.282^{+0.010}_{-0.009}$ & 
$0.285_{-0.010}^{+0.012}$  & $0.696_{-0.037}^{+0.025}$ & $3.36_{-2.16}^{+\infty}$  & 572.07 / 631 \rule{0pt}{1.2em}  \\
 \hline
$\!F(R)\;$(\ref{FR1})$\!$& $\chi^2_{tot}$& $0.280_{-0.001}^{+0.001}$ &
$0.294_{-0.007}^{+0.009}$  & $0.637_{-0.062}^{+0.047}$ & $2.38_{-0.80}^{+\infty}$  & 575.51  / 634 \rule{0pt}{1.2em} \\
  \hline
$\!\Lambda$CDM$\!$ & $\chi^2_{\Sigma3}$& $=\Omega_m^0$ & 
$0.282^{+0.010}_{-0.009}$  & $0.718_{-0.010}^{+0.009}$ & $\infty$  & 572.93 / 633 \rule{0pt}{1.2em} \\
 \hline
$\!\Lambda$CDM$\!$ & $\chi^2_{tot}$& $=\Omega_m^0$ &
$\!0.2772_{-0.0004}^{+0.0003}\!$  & $\!0.7228_{-0.0003}^{+0.0004}\!$ & $\infty$  & 583.24 / 636 \rule{0pt}{1.2em} \\
 \hline \end{tabular}
\caption{Predictions of the exponential $F(R)$ model (\ref{FR1}) and the $\Lambda$CDM
for different data sets ($\chi^2_{\Sigma3}=\chi^2_{SN}+\chi^2_H+\chi^2_{BAO}$,
$\chi^2_{tot}=\chi^2_{\Sigma3}+\chi^2_{CMB}$): $\min\chi^2$ and
  $1\sigma$ estimates of model parameters.}
 \label{Estim}}\end{table}


\begin{figure}[th]
   \centerline{ \includegraphics[scale=0.68,trim=5mm 0mm 2mm -1mm]{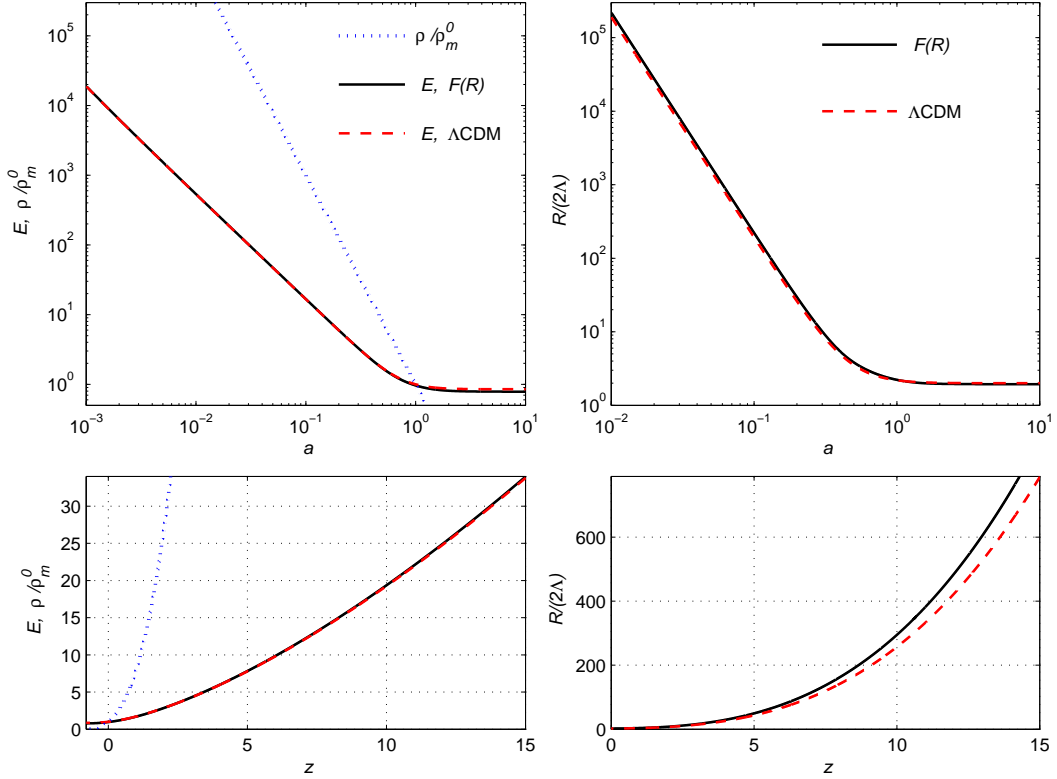}}
\caption{  The top panels in the logarithmic scale illustrate the dependence of 
$E=H/H_0^{\Lambda  CDM}$,  $\rho/\rho_m^0$ (the blue dots) and ${\cal R}=R/(2\Lambda)$
on $a$ for the $F(R)$ model (\ref{FR1}) (the black solid lines) and for the $\Lambda$CDM
(the red dashed lines). The corresponding plots $E(z)$,  $\rho(z)/\rho_m^0$ and ${\cal
R}(z)$ in the usual scale are shown at the bottom panels. For both models the parameters
are from Table~\ref{Estim} for $\chi^2_{tot}$.} 
  \label{Fphase}
\end{figure}

In Fig.~\ref{Fphase} we demonstrate how the numerical solution of the system
(\ref{eqE2}), (\ref{eqr2}) behaves (the black solid lines in the phase space diagram) in
comparison with the $\Lambda$CDM model (the red dashed lines). We show the plots for the
Hubble parameter $E=H/H_0^{\Lambda  CDM}$ and the Ricci scalar ${\cal R}=R/(2\Lambda)$
(in the right panels) depending on  $a$ in the top panels with the logarithmic scale and
the same plots $E(z)$, ${\cal R}(z)$ (with $z$ instead of $a$) in the usual scale in the
bottom panels. The model parameters for both models are taken from Table~\ref{Estim},
they are optimal for $\chi^2_{tot}$.

One can see that for the optimal parameters the $F(R)$ and $\Lambda$CDM models
demonstrate rather close dynamics of the Hubble parameter $E$ with the small future
divergence for $a>1$. The plots for the density $\rho/\rho_m^0$ (the blue dots) are the
same for both models.

\begin{figure}[th]
   \centerline{ \includegraphics[scale=0.6,trim=5mm 0mm 2mm -1mm]{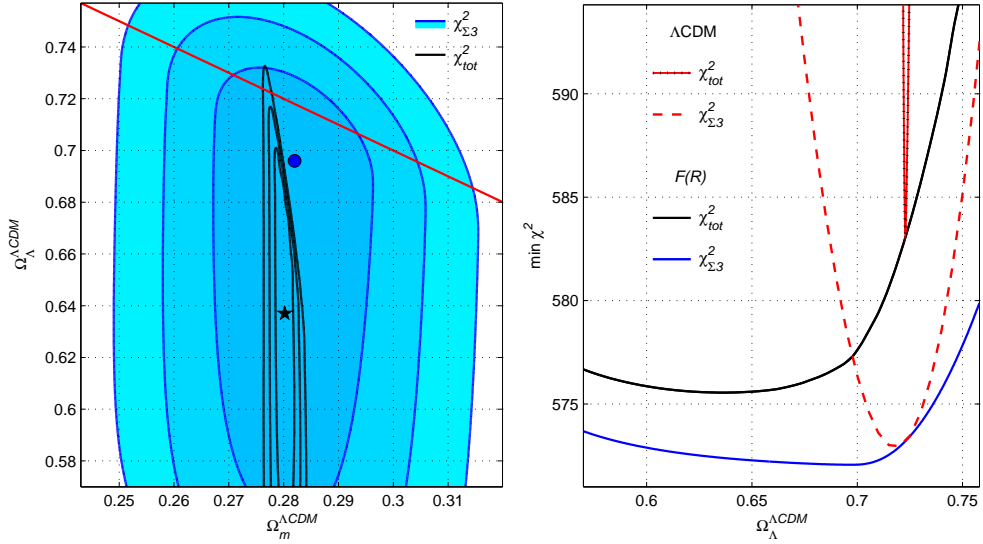}}
\caption{Contour plot for the $\Omega_m^{\Lambda CDM}-\Omega_\Lambda^{\Lambda CDM}$ plane and the one-dimensional distribution $\chi^2(\Omega_\Lambda^{\Lambda CDM})$ for $\chi^2_{tot}$ (solid black
lines) and for $\chi^2_{\Sigma3}$ (solid blue lines and filled contours). The red lines correspond to the $\Lambda$CDM model.}
  \label{F3}
\end{figure}

For the parameter $\Omega_\Lambda^{\Lambda CDM}$, the optimal values and $1\sigma$
errors in Table~\ref{Estim} were found after preliminary calculation of two-dimensional
distributions $\chi^2(\Omega_m^{\Lambda CDM},\Omega_\Lambda^{\Lambda CDM})$ for
$\chi^2_{\Sigma3}$ and $\chi^2_{tot}$. The contour plots for these two-dimensional
distributions and the corresponding one-dimensional plot for
$\chi_{min}^2(\Omega_\Lambda^{\Lambda CDM})=\min\limits_{\beta,\Omega_m^{\Lambda
CDM}}\chi^2$ are depicted in Fig.~\ref{F3} with the same notations. We see that the CMB
data (\ref{CMB}-\ref{CMBpriors}) for $\chi^2_{tot}$ essentially restrict the $1\sigma$
bounds of $\Omega_m^{\Lambda CDM}$ in comparison with the $\chi^2_{\Sigma3}$ function
for SNe${}+H(z)+{}$BAO data. This is also connected with the factor $\sqrt{\Omega_m^0}$
in the parameter $R$ (\ref{CMB}) and the narrow restrictions (\ref{CMBpriors}). The
minimum value for $\chi^2_{tot}(\Omega_m^{\Lambda CDM},\Omega_\Lambda^{\Lambda CDM})$
(the black star in Fig.~\ref{F3}) is shifted lower from the ``$\Lambda$CDM line''
$\Omega_\Lambda^{\Lambda CDM}+\Omega_m^{\Lambda CDM}=1$.

\begin{figure}[th]
   \centerline{ \includegraphics[scale=0.66,trim=5mm 0mm 2mm -1mm]{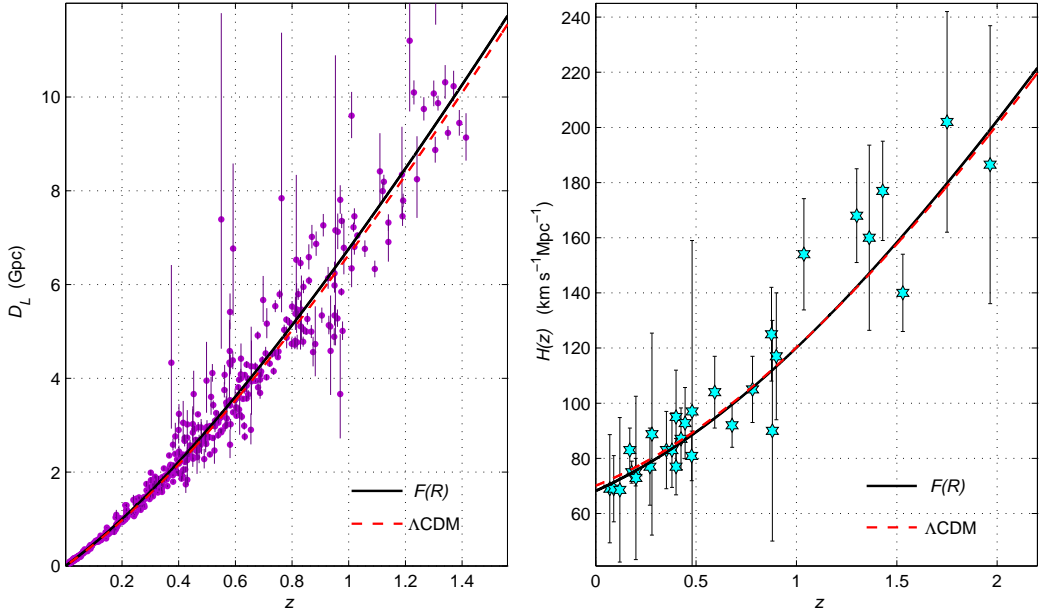}}
\caption{Luminosity distance (left panel) and Hubble parameter (right panel) for the best fit when considering the general case described by (\ref{OmmL}) for the model (\ref{FR1}). The best fit for $\Lambda$CDM is also depicted.}
  \label{F4}
\end{figure}

In the right panel of Fig.~\ref{F3} we compare the predictions of the $F(R)$ model
(\ref{FR1}) and the $\Lambda$CDM model, where $\Omega_\Lambda^{\Lambda
CDM}\equiv\Omega_\Lambda^0=1-\Omega_m^0$. The  $\Lambda$CDM dependencies
$\chi^2(\Omega_\Lambda^0)$ coincide with their analogs for the  model (\ref{FR1}) under
the restriction (\ref{OmmL1}) $\Omega_m^{\Lambda CDM}+\Omega_\Lambda^{\Lambda CDM}=1$
(after minimising over $\beta$). In the  $\Lambda$CDM case, the $1\sigma$ errors are
essentially smaller. Finally, Fig.~\ref{F4} shows the evolution of the luminosity
distance and the Hubble rate for the exponential gravity model and for $\Lambda$CDM for
their best fits. As shown, both curves fit the data similarly, such that both models
become indistinguishable.

\section{Exponential model and inflation}
\label{Inflation}

The exponential $F(R)$ model (\ref{FR1}) considered in the previous sections, describes
all observational manifestations of the late-time acceleration. 
However, such a model can also explain the early-time inflation when introducing some suitable modifications in the form of $F(R)$ as follows
\cite{ElizaldeNOSZ11}:
\begin{equation}
F(R)= R-2\Lambda\bigg[1-\exp\Big(-\beta\frac{R}{2\Lambda}\Big)\bigg]-
\Lambda_i\bigg[1-\exp\bigg(-\Big(\frac R{R_i}\Big)^n\bigg)\bigg]+\gamma R^\alpha.
 \label{FR2}\end{equation}
These additional terms can generate the expected cosmological constant $\Lambda_i$
 during the inflationary era, when $R$ is near or larger than $R_i$. The natural
number $n > 1$ helps to avoid the effects of inflation during the matter era when $R\ll
R_i$ and the last term $\gamma R^\alpha$ in Eq.~(\ref{FR2}) is necessary for a
successful exit from inflation. As pointed out in Ref.~\cite{ElizaldeNOSZ11}, the model
(\ref{FR2}) has the following properties: a de Sitter phase naturally arises during
inflation in the high-curvature regime, the inflationary terms become negligible in the
small curvature era $R\ll R_i$, inflation ends successfully and avoids anti-gravity
effects and instabilities during the matter era. To satisfy all these properties, the
following  requirements are obtained over the free parameters:
  \be 2<\alpha<3\ , \qquad
n>\alpha\ , \qquad R_i=2\Lambda_i,\qquad \gamma\simeq\Lambda_i^{1-\alpha}\ .
\label{conditions1} \ee
 where $\Lambda_i$ can vary in the range
\begin{equation}
\frac{\Lambda_i}{\Lambda}=10^{86}-10^{104}.
 \label{Lambi}\end{equation}

In addition, inflation occurs when $R\gg R_i$, where an unstable de Sitter point $R=
R_{dS}$ arises in the phase space, if the condition  \cite{ElizaldeNOSZ11}
 \begin{equation}
G(R_{dS})=0
 \label{Geq0}\end{equation}
for the function
 $$
 G(R)=2F(R)-RF_R
 $$
 is satisfied. The condition (\ref{Geq0}) may be deduced, if we search a de Sitter solution $R=
R_{dS}={}$const, $H={}$const of the system (\ref{dynam}) before the matter era.

If $R_{dS}/R_i>1.5$, we can neglect the $e^{-(R/R_i)^n}$ terms in $G(R)$ and the
 condition (\ref{Geq0}) for de Sitter solutions takes the form
 \be
 R_{dS}-(\alpha-2)\gamma R_{dS}^\alpha-2\Lambda_i=0\ ,
 \label{Condi11} 
 \ee
Here the constant $\gamma$ in Eq.~(\ref{FR2}) is expressed as
 $\gamma=(\Gamma\Lambda_i)^{1-\alpha}$. In Ref.~\cite{ElizaldeNOSZ11} the de Sitter solution with parameters $n=4$,
$\gamma=(4\Lambda_i)^{1-\alpha}$, $\alpha=5/2$, $R_{dS}=4\Lambda_i$ was considered. The
corresponding values here are $\Gamma=4$, $R_{dS}=\Gamma\Lambda_i$ with $\alpha=5/2$ satisfy the
condition (\ref{Condi11}). However, below we analyze a more wide set of inflationary
solutions which obey the observational limitations for the slow-roll parameters.\\

Let us now focus on the realisation of slow-roll inflation within the model (\ref{FR2})
and its predictions. In order to do so, the scalar-tensor counterpart of $F(R)$
gravities is more convenient than its original action, such that the $F(R)$ gravities
can be expressed in terms of a scalar field as shown in (\ref{Act2}) and (\ref{dsg2}).
By varying the action (\ref{Act2}) with respect to the scalar field $\phi$, it yields
  \be R-\frac{\partial V(\phi)}{\partial
\phi}=0\quad \rightarrow \quad \phi=\phi(R)\quad \rightarrow \quad F(R)=\phi(R)
R-V(\phi(R)) . \label{scalar2} \ee
 Here recall that the relations  (\ref{dsg2}) for the scalar field and its potential
hold:
 $$
 \phi=F_{R}\ , \qquad V(\phi)=RF_R-F.$$
  In order to
analyse  slow-roll inflation for the model (\ref{FR2}), the action (\ref{Act2}) can be
transformed into the Einstein frame by the following conformal transformation:
 \be
\tilde{g}_{\mu\nu}=\Omega^2 g_{\mu\nu}\ \quad \text{where} \quad \Omega^2=\phi\ ,
\label{scalar4} \ee
 And the action (\ref{Act2}) turns out:
 \be \tilde{S}=\int
{\rm
d}^4x\sqrt{-\tilde{g}}\left[\frac{\tilde{R}}{2\kappa^2}-\frac{1}{2}\partial_{\mu}\tilde{\phi}\,\partial^{\mu}\tilde{\phi}-\tilde{V}(\tilde{\phi})\right]\
. \label{scalar5}
 \ee
  Here, we have redefined the scalar field to keep the kinetic term
in a canonical form:
 \be \phi=\e^{\sqrt{\frac{2}{3}}\kappa\tilde{\phi}}\ , \quad
\tilde{V}=\frac{\e^{-2\sqrt{\frac{2}{3}}\kappa\tilde{\phi}}}{2\kappa^2} V\ .
\label{scalar6} \ee

The FLRW equations for the action (\ref{scalar5}) become simpler than working directly
on the $F(R)$ action:
\bea \frac{3}{\kappa^2} H^2 = \frac{1}{2}\dot{\tilde{\phi}}^2 +
\tilde{V}(\tilde{\phi})\, ,  \nn - \frac{1}{\kappa^2} \left( 3 H^2 + 2\dot H \right)=
\frac{1}{2}\dot{\tilde{\phi}}^2 - \tilde{V}(\tilde{\phi})\, , \label{scalar7} \eea
 whereas the scalar field equation is given by:
 \be
\ddot{\tilde{\phi}}+3H\dot{\tilde{\phi}}+V'(\tilde{\phi})=0\ . \label{scalar8} \ee
During slow-roll inflation the scalar field mimics an effective cosmological constant,
what basically means that $H\dot{\tilde{\phi}}\gg \ddot{\tilde{\phi}}$ and $\tilde{V}\gg
\dot{\tilde{\phi}}^2$. Both conditions can be also expressed through the so-called
slow-roll parameters
 \be \epsilon= \frac{1}{2\kappa^2} \left(
\frac{\tilde{V}'(\tilde{\phi})}{\tilde{V}(\tilde{\phi})} \right)^2\, ,\qquad \eta=
\frac{1}{\kappa^2} \frac{\tilde{V}''(\tilde{\phi})}{\tilde{V}(\tilde{\phi})}\,.
\label{scalar9} \ee
 These quantities remain very small when inflation occurs such that
$\epsilon\ll 1$ and $\eta<1$, while at the end of inflation $\epsilon\gtrsim 1$. In
addition, the slow-roll parameters (\ref{scalar9}) are related to the amplitude and
scale dependence of the perturbations originated during inflation, such that the
spectral index of the perturbations and the tensor-to-scalar ratio, are given by:
  \be
n_\mathrm{s} - 1= - 6 \epsilon + 2 \eta\, ,\qquad r = 16 \epsilon \ . \label{scalar10}
\ee
 Since both values are very well constrained by the last data from Planck and Bicep2
collaborations \cite{Planck-Inflation}, which give:
  \be n_\mathrm{s}=0.968\pm0.006\ ,\qquad
r<0.07\ . \label{Planck} \ee

Then, we can test whether the model (\ref{FR2}) is capable
to satisfy such constraints. Firstly, let us analyse the action (\ref{FR2}), as during
inflation $R\gg \Lambda_i$, the action (\ref{FR2}) can be approximated as follows:
\begin{equation}
F(R)\sim R-\Lambda_i +\gamma R^\alpha\ .
 \label{FR2bis}\end{equation}
Then, by the expression in (\ref{scalar6}), the relation among the scalar field and the curvature is obtained:
\be
R=\Gamma\Lambda_i\left(-\frac{1-\e^{\sqrt{\frac{2}{3}}\kappa\tilde{\phi}}}{\alpha}\right)^{\frac{1}{\alpha-1}}\ .
\label{scalar11}
\ee
Here recall that $\gamma=(\Gamma\Lambda_i)^{1-\alpha}$. During inflation $R\gg \Lambda_i$ and consequently $\kappa\tilde{\phi}\gg 1$, so the above expression (\ref{scalar11}) can be approximated by:
\be
R\sim\Gamma\Lambda_i\left(\frac{\e^{\sqrt{\frac{2}{3}}\kappa\tilde{\phi}}}{\alpha}\right)^{\frac{1}{\alpha-1}}\ .
\label{scalar12}
\ee
While the scalar potential yields:
\be
\tilde{V}(\tilde{\phi})=\frac{\Lambda_i}{2\kappa^2}\frac{1+\Gamma(\alpha-1)\left(\frac{\e^{\sqrt{\frac{2}{3}}\kappa\tilde{\phi}}}{\alpha}\right)^{\frac{\alpha}{\alpha-1}}}
{\e^{2\sqrt{\frac{2}{3}}\kappa\tilde{\phi}}}\ .
\label{scalar13}
\ee
Inflation usually requires a number of e-foldings $N \simeq 55 - 65$, which is defined as
\begin{align}
N  \equiv  \int_{t_{start}}^{t_{end}} \tilde{H} {\rm d}t.
\label{scalar14}
\end{align}
Applying the slow--roll approximation to the above relation (\ref{scalar14}), the number of e-foldings yields:
\be
N \simeq -\kappa^2 \int_{\tilde{\phi}_{start}}^{\tilde{\phi}_{end}} \frac{\tilde{V}(\tilde{\phi})}{\tilde{V}'(\tilde{\phi})} {\rm d}\phi\ ,
\label{Nefolds}
\ee
where $\tilde{\phi}_{start}>>\tilde{\phi}_{end}$. For the model analysed here, this expression can not be solved analytically even by taking some approximations. Hence, in order to illustrate the powerful of the model, let us consider an example for the parameter $\alpha$ of the model that satisfy the conditions (\ref{conditions1}):
\be
\alpha=2.001\ , \quad \Gamma=2\ .
\label{scalar15}
\ee
Note that the value $\alpha=2$ corresponds to Starobisnky inflation. Then, by integrating (\ref{Nefolds}), the following results are obtained:
\be
N\sim58\ , \quad n_s=0.965\ ,\quad r=0.0034\ .
\label{scalar16}
\ee
As shown above in (\ref{Planck}), these values lie within the allowed ranges provided by Planck, such that the model (\ref{FR2}) can reproduce well inflation and then recover late-time acceleration, leading to a unified description of the universe evolution. \\

Let us now analyse the the system  of equations (\ref{dynam}) during the inflationary epoch, which
are reduced to
\begin{eqnarray}
\frac{d\log E}{dx}&=&\Omega_\Lambda^{\Lambda CDM}\frac{{\cal R}}{E^2}-2, \nonumber \\   
\frac{d\log {\cal R}}{dx}&=&\frac{\Omega_\Lambda^{\Lambda CDM}\big[\lambda_i(1-e_i)+2r_i^3{\cal R}e_i
+(\alpha-1)\tilde\gamma{\cal R}^\alpha\big]/E^2-1+2r_i^3e_i
-\alpha\tilde\gamma{\cal R}^{\alpha-1}}{{\cal R}\big[\lambda_i^{-1}(4r_i^6-3r_i^2)e_i
+\alpha(\alpha-1)\tilde\gamma{\cal R}^{\alpha-2}\big]}.
  \nonumber \end{eqnarray}
Here $\lambda_i=\Lambda_i/(2\Lambda)$, $r_i=R/R_i$, $e_i=e^{-r_i^4}$, $\tilde\gamma=(\Gamma\lambda_i)^{1-\alpha}$.

\begin{figure}[th]
   \centerline{ \includegraphics[scale=0.75,trim=5mm 0mm 2mm -1mm]{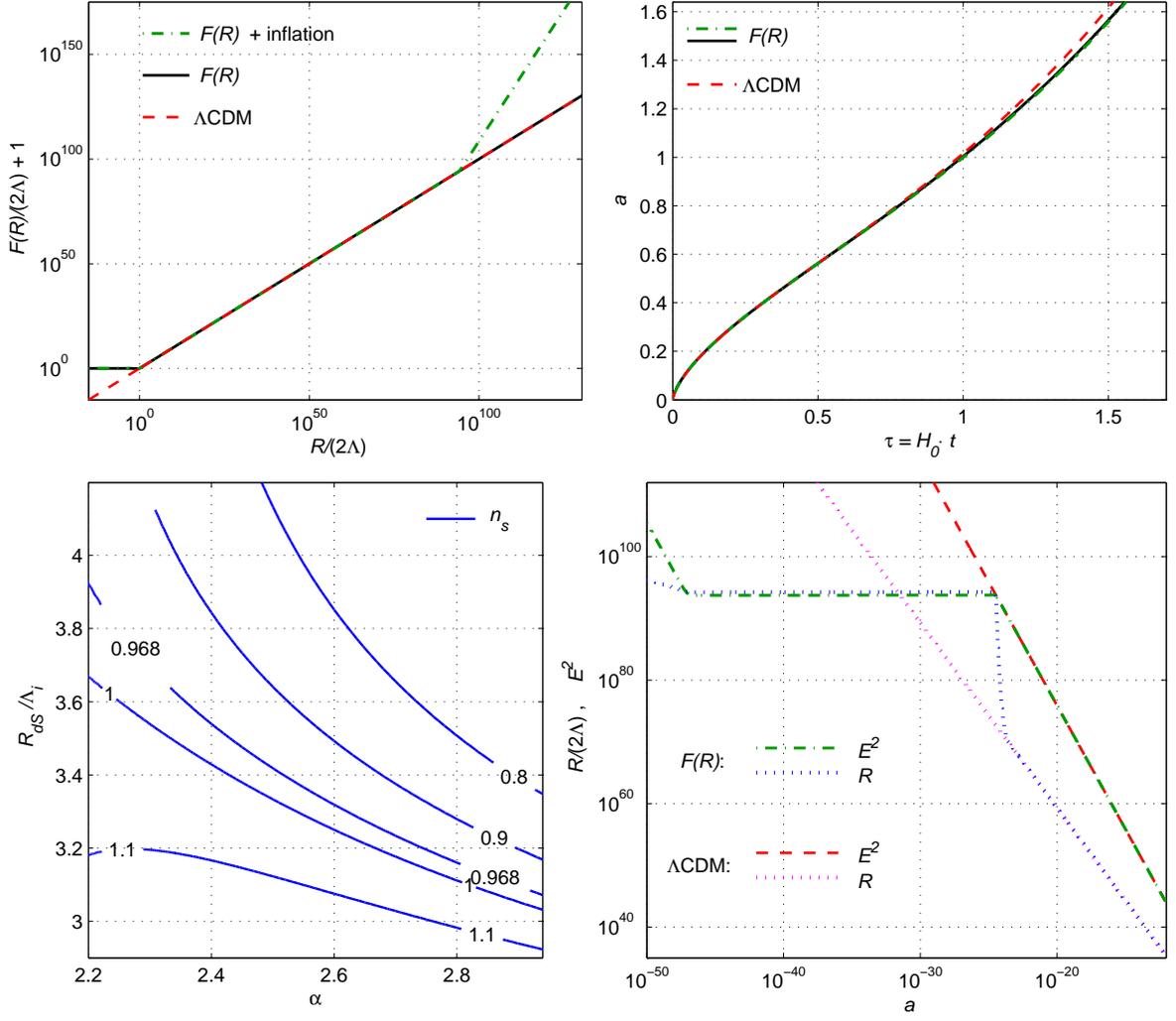}}
\caption{Plots of $F(R)$ (top-left), $a(\tau)$ (top-right), $E^2(a)$ and ${\cal R}(a)$ (the
bottom-right panel) for the models: (\ref{FR1}), (\ref{FR2}) and the $\Lambda$CDM model with the parameters from Table
\ref{Estim}.  In the bottom-left panel the level lines of $n_\mathrm{s}$
are shown for the model (\ref{FR2}). }
  \label{F5}
\end{figure}

In the top-left panel of Fig.~\ref{F5}, the $F(R)$ function (\ref{FR2}) is depicted for
the values  
\be
n=4\,,\quad \alpha=5/2\,,\quad \psi=0.883\,,\quad\Gamma=3.871\,,\quad R_{dS}/\Lambda_i=3.419\,,
\quad \Lambda_i/\Lambda=10^{94}
\,, \label{optim} \ee
and compared to the exponential gravity (\ref{FR1}) and the $\Lambda$CDM model. As shown, both $F(R)$ models
are not distinguishable at low curvature regimes but they are when the curvature becomes
very large, as during inflation. Moreover, the model  (\ref{FR2}) practically coincides
with  the $\Lambda$CDM model in the range $\Lambda<R<R_i$ (or $1<{\cal
R}<\Lambda_i/\Lambda$), while differs for $R>R_i$ because of the $\gamma R^\alpha$ term.
Finally for $R\sim H_0^2$, both $F(R)$ models (\ref{FR1}) and (\ref{FR2}) behave
similarly and deviate from $\Lambda$CDM.

One can conclude that the difference between the models  (\ref{FR1}) and (\ref{FR2}) is
negligible during the radiation/matter era, so their observational predictions
coincide. In order to illustrate such behaviour for both $F(R)$ models as well as
for the $\Lambda$CDM model, the top-right panel
in Fig.~\ref{F5} depict the evolution of the scale factor  $a(\tau)$, where
$\tau=H_0t$ is dimensionless time. Here the differences  between the $F(R)$ models
are not visible, because it takes place for $a\ll10^{-20}$, but the late time and future
behaviour of  the $\Lambda$CDM model differs from that for the $F(R)$ models.

The early time evolution and the inflationary epoch ($R\gg R_i$ and $R\simeq R_i$)
of the model  (\ref{FR2} in comparison with the $\Lambda$CDM model  are shown in the bottom-right panel
of Fig.~\ref{F5}. Here the dashed and dash-dotted lines correspond to
the Hubble parameter  $E^2(a)$ and dots describe the Ricci scalar ${\cal R}(a)$
in their normalized form (\ref{Er})  as functions of the scale factor.

The  inflationary solution is unstable \cite{ElizaldeNOSZ11}: after $N \simeq 55$ e-foldings
the de Sitter solution decays and the evolution transforms into the $\Lambda$CDM  behaviour.

\section{Conclusions}
\label{conclusions}

Exponential gravity may be considered as an alternative to the so-called concordance model. The main gravitational action studied along this manuscript and described in (\ref{FR1}) represents an slightly correction to the usual Hilbert-Einstein action with a cosmological constant. Such correction is modelled by a free parameter, which has been called $\beta$, such that controls the scale at which corrections to GR become important. As shown in some previous works \cite{CognolaENOSZ08,ElizaldeNOSZ11}, such type of $F(R)$ models can reproduce a late-time acceleration epoch. However, the aim of this work was to show in an accurate way, whether such type of gravitational actions fulfil the necessary cosmological constraints leading to a reasonable bound on the crucial parameter $\beta$. Note that an equivalent analysis was performed in \cite{YangLeeLG2010} and more recently in \cite{ChenGLLZ14}. In our paper we have updated the constraints obtained in previous works by assuming new released data from the last years. In addition, we have also analysed the exponential model at late-times by following two approaches: the first one by assuming the condition (\ref{OmmL1}) and the second by assuming a more general approach. The former provides a more restricted case as we are forcing the model to mimic $\Lambda$CDM at the present time, while the latter keeps the model free, except for the initial conditions which are the same for both cases. As expected, the more restrictive on the theoretical conditions are, the better the constraints on the free parameter turn out, as shown in the bottom panels of Fig.~\ref{F2}. On the other hand, whereas the $\beta$ parameter is not upper bounded (recall that GR is recovered for $\beta\rightarrow\infty$), the fits realised in the paper, where SNe Ia, BAO, CMB and $H(z)$ data were used, provides a sufficient small lower bound on $\beta$ that may have consequences at the perturbation level, an aspect that should be studied in future. In addition, the values for the matter density $\Omega_m$ do not differ too much among the one given by the $\Lambda$CDM model and the one from the exponential gravity model, either when some approximations are assumed or when the general case is considered. Moreover, the $\chi_{min}^2$ is a bit smaller for the exponential gravity case than in $\Lambda$CDM, such that the best fit does not correspond to $\Lambda$CDM, although the difference is not statistically significant. From a qualitative point of view, the tiny differences among exponential gravity  and $\Lambda$CDM can be shown by looking at the form of the action in (\ref{FR1}). By a sufficient small exponent, i.e. a $\beta$ parameter large enough, the Lagrangian mimics quite well the Hilbert-Einstein action with a cosmological constant, such that there is not significant difference on the cosmological evolution among both models, as shown in Fig.~\ref{F3}.  \\

In addition, the exponential gravity action (\ref{FR1}) can be extended in such a way that the new action (\ref{FR2}) can reproduce inflation as well. Note that inflationary models within the framework of $F(R)$ theories have been widely studied in the literature, as shown by one of the most successful inflationary models, the so-called Starobinsky inflation \cite{Starobinsky:1980te}. In this sense, we have proposed here a model where the exponential term responsible for the late-time acceleration is suppressed at the large curvature regime and consequently its induced cosmological constant while two additional terms may become important: a different effective cosmological constant $\Lambda_i$ (much larger than $\Lambda$) and a power term $R^{\alpha}$. Recalling that Starobinsky inflation is described by a $R^2$ term, the inclusion of $R^{\alpha}$ just generalised the latter and ensures a successful exit from inflation \cite{ElizaldeNOSZ11}. However, as shown in some previous papers, such exponent has to be $2<\alpha<3$ in order to avoid instabilities \cite{ElizaldeNOSZ11}. Here, we have extended the previous analysis by using the usual scalar-tensor counterpart and obtaining the explicit form of the potential for the scalaron. Then, we computed explicitly the spectral index of curvature fluctuations during inflation. An example fulfilling all the requirements provides an spectral index that leads to a nearly invariant power spectrum and an almost null amplitude for the tensor modes, predictions in agreement with the last data released by the Planck collaboration. Note that the constant $\Lambda_i$ establishes the energy scale at which the last terms in the action (\ref{FR2}) become important, such that then the action also provides a quasi-de Sitter inflationary expansion, similar to Starobinsky model, where the $R^{\alpha}$ term guarantees a successful exit from inflation. As shown, the values for the free parameters which provide the correct values for the spectral index and the scalar-to-tensor ratio, also avoid further corrections when inflation ends. Such model is then able to reproduce inflation and successfully exit.   \\

Hence, we can conclude that the full gravitational Lagrangian (\ref{FR2}) is capable of reproducing inflation and late-time acceleration in such a way that no other fields are required. Recently, one more extension of this type of exponential gravity with log-corrected $R^2$ term was proposed to explain the unified universe history (see Ref.~\cite{Odintsov:2017hbk}). As shown here, the Lagrangian satisfies the observational constraints, with a no statistical significant difference with respect to the $\Lambda$CDM model, what means that they can not be distinguished. Simultaneously, the $F(R)$ model (\ref{FR2}) provides the right predictions for the inflationary epoch. Next steps should be focused on the analysis of cosmological perturbations and the possible effects of such Lagrangian at the astrophysical level.

\section*{Acknowledgements}
SDO and DSG acknowledge the support by MINECO (Spain), project FIS2013-44881, FIS2016-76363-P and by CSIC I-LINK1019.
DSG is funded by the Juan de la Cierva program (Spain) No.~IJCI-2014-21733. This article is based upon work from CANTATA COST (European Cooperation in Science and Technology) action CA15117,  EU Framework Programme Horizon 2020.

\end{document}